\documentclass[12pt]{article}
\usepackage{geometry}
\usepackage{color}
\usepackage{xcolor}
\usepackage{a4}
\usepackage{graphicx}
\usepackage{epsf}
\usepackage{amsmath}
\usepackage{amssymb}
\usepackage{cite}
\newcommand{\be}{\begin{equation}}
\newcommand{\ee}{\end{equation}}

\newcommand{\Rmnum}[1]{\expandafter\@slowromancap\romannumeral #1@}
\newcommand{\bea}{\begin{eqnarray}}
\newcommand{\eea}{\end{eqnarray}}
\begin{document}
\def\C{{\mathbb{C}}}
\def\R{{\mathbb{R}}}
\def\s{{\mathbb{S}}}
\def\T{{\mathbb{T}}}
\def\Z{{\mathbb{Z}}}
\def\W{{\mathbb{W}}}
\def\Bbb{\mathbb}
\def\BZ{\Bbb Z} \def\BR{\Bbb R}
\def\BW{\Bbb W}
\def\BM{\Bbb M}
\def\BC{\Bbb C} \def\BP{\Bbb P}
\def\CP{\BC\BP}
\begin{titlepage}
\title{Astrophysics of Bertrand Space-times} \author{} 
\date{
Dipanjan Dey, Kaushik Bhattacharya, Tapobrata Sarkar 
\thanks{\noindent 
E-mail:~ deydip, kaushikb, tapo @iitk.ac.in} 
\vskip0.4cm 
{\sl Department of Physics, \\ 
Indian Institute of Technology,\\ 
Kanpur 208016, \\ 
India}} 
\maketitle 
 
\abstract{We construct a model for galactic dark matter that arises as a solution of Einstein gravity, and is a Bertrand space-time matched with an
external Schwarzschild metric. This model can explain galactic rotation curves. Further, we study gravitational lensing in these 
space-times, and in particular we consider Einstein rings, using the strong lensing formalism of Virbhadra and Ellis. Our results are in good agreement with observational
data, and indicate that under certain conditions, gravitational lensing effects from galactic dark matter may be similar to that from Schwarzschild backgrounds.} 

\end{titlepage}

\section{Introduction}
\label{intro}

Over the last few decades, several models for galactic dark matter have been proposed, which seek to explain astrophysical observations. 
For example, the popular Navarro-Frenk-White (NFW) model \cite{nfwprof} provides excellent fits to observational data for a wide range of galaxies. 
Galactic dark matter models such as the NFW model have mostly been constructed via numerical techniques. A natural question that arises in this context is the
possible role of General Relativity (GR) in galactic astrophysics, i.e is there a space-time metric that can explain experimental observations ? 
Although several works in the literature (see, e.g. \cite{sayan}, \cite{mdroberts}) deal with the effects of GR in galactic dynamics and construct galactic metrics, 
it is fair to say that this approach has not been very popular. Along with various subtleties in GR regarding observers and co-ordinate frames, it  
is also generally believed that GR effects at the galactic scale would possibly be small compared to those of local dynamics. However, it is
not difficult to imagine that GR might play an important role in the dynamics of stars, by seeding the space-time in which they move, via a matter distribution \cite{dbs1}. 

Imagine a star in a galaxy dominated by dark matter. Its motion, via the Equivalence
Principle, can, to a good approximation, be thought of as a time-like geodesic, in a space-time that is sourced by galactic dark matter.  In
other words, the star is in free fall with respect to all other neighboring heavenly objects, in a dark matter dominated
background. Observations indicate that typically stars that are far from the galactic centers move in approximately closed circular
trajectories \cite{binneytremaine}. The question thus reduces to whether in the GR framework, one can describe such a motion, or, in
other words, construct a space-time in which stable closed orbits are allowed at each point in space. Such a problem was envisaged and
solved by Perlick more than two decades back \cite{perlick}. Perlick's work shows that it is possible to define
Bertrand space-times (BSTs), \footnote{There are two distinct classes of BSTs discovered by Perlick, which are called BSTs of Type I and
Type II. What we deal with in this paper is a class of BSTs of Type II. We will, for ease of presentation, refer to these as BSTs, throughout this paper.}
which are solutions of Einstein's equations, and have the property that every point in a spatial
hypersurface admits closed, stable orbits.\footnote{This generalizes Bertrand's theorem in classical mechanics
\cite{bert},\cite{goldstein} to a GR scenario (for related work in the special relativistic case, see \cite{kb}).} This is precisely
the property that a space-time seeded by galactic dark matter should have, at least in the disc or halo regions, which make BSTs
interesting objects to study. Of course, any realistic space-time model must satisfy the energy conditions associated with GR, and in \cite{kbs1}, \cite{kbs2} 
it was shown that a simple class of BSTs derivable from Perlick's original work do satisfy the weak and strong energy conditions.

In \cite{dbs1}, we showed that the matter seeding the BSTs can be interpreted as galactic dark matter, 
\footnote{Our results are maximally applicable to low surface brightness (LSB) galaxies.} i.e it supports flat rotation
curves. \footnote{For a different perspective on galactic rotation curves and dark matter, see e.g. \cite{salucci}.}
It was also shown that near the knee regions of these rotation curves, a Newtonian approximation results in 
an analytical derivation of an NFW profile, while away from this region, a Hernquist \cite{hernquist} profile is predicted. This approximation also gave good
estimates of galactic masses, after fitting the BST parameters with observed rotation curves. It was further established that the dark matter that sources 
BSTs can be described by an anisotropic two-fluid model. The radial distance where the fluid pressure in such models becomes negligible gave a natural 
estimate of the size of the galaxies. 

One of our main motivations in this paper is to further justify BSTs as possible models for galactic dark matter. To this end, we construct a solution 
of Einstein's equations, which contain an internal BST and an external vacuum Schwarzschild space-time. Since BSTs are not asymptotically flat, this construction 
is necessitated on two counts. Firstly, it gives us insight into the possible interpretation of rotation curves in GR and helps us to compare BSTs with some 
other popular metrics used in the literature. Further, it allows us to calculate gravitation 
lensing in BSTs, using the formalism of Virbhadra and Ellis  \cite{ve1}, \cite{ve2} (for closely related work, see \cite{v1}, \cite{v2}, \cite{v3}. For related formalisms, 
see, e.g \cite{bozza1}, \cite{bozza2}). We study rotation curves and gravitational lensing in our model, and our main conclusion is that BSTs provide a viable 
general relativistic model of galactic dark matter, as our results are in good agreement with observational data. We are also able to calculate gravitational lensing
from a galactic cluster, using our formalism. Our results indicate that as far as gravitational lensing is concerned, galactic dark matter might behave similar 
to Schwarzschild backgrounds. 

This paper is organized as follows. In section 2, after reviewing some general features of BSTs, we compare them with some other popular space-time models. 
In section 3, we construct a solution of Einstein's equation that is internally a BST and externally Schwarzschild. In particular, the Darmois - Israel junction conditions 
are studied, and we show how this conditions can be satisfied in our case. In Section 4, studies on gravitational lensing properties of our
solution is presented and compared with observations. Section 5 ends the paper with some discussions
and directions for future research.

\section{BSTs and some relevant space-times with singularities}

To define a Bertrand space-time, one starts with a static, spherically symmetric Lorentzian manifold, whose metric, in
terms of two functions $\lambda(r)$ and $\nu(r)$ take the form 
\begin{eqnarray}
ds^2 = -e^{2\nu(r)} dt^2 + e^{2\lambda(r)} dr^2 + r^2 (d \theta^2 +
\sin^2 \theta d \phi^2).
\label{invl}
\end{eqnarray}
Here, $r$ is a radial co-ordinate, $t$ the time, and $\theta$, $\phi$ are standard co-ordinates on the
two-sphere. \footnote{We generally use units where the velocity of light $c=1$. Wherever required, the units are restored.}
Such a Lorentzian manifold is called a BST, provided there is a closed circular orbit passing through each point, which is 
stable under small perturbations of initial conditions.  From this definition, Perlick \cite{perlick} showed that there can be two distinct 
families of BSTs. Here, we will deal with one of them (called BSTs of type II by Perlick) given by the metric 
\begin{eqnarray}
ds_{\rm BST}^2= -\frac{dt^2}{D + \alpha\sqrt{r^{-2}+K}} + \frac{dr^2}
{\beta^2(1+Kr^2)}+r^2(d\theta^2 + \sin^2 \theta\,d\phi^2)\,. 
\label{type2}
\end{eqnarray}
The parameters $D$, $K$, $\alpha$ are real, and $\beta$ must be a positive rational number. Here $D$ and $\beta$ are dimensionless
constants whereas $\alpha$ has the dimension of length and $K$ has the dimension of inverse squared length. In order to construct a
phenomenological model of a galactic space-time using the metric of Eq.~(\ref{type2}), we minimally require three parameters, one fixing a
galactic scale, a second to parameterize the nature of closed orbits, and a third to fix the velocity in such an orbit. Such a space-time can
be specified as:
\begin{eqnarray}
ds_{\rm BST}^2 = -\frac{dt^2}{D +\frac{\alpha}{r}} + \frac{dr^2}{\beta^2}
+r^2(d\theta^2 + \sin^2 \theta\,d\phi^2)\,.  
\label{type2a}
\end{eqnarray}
where we have set $K=0$ in Eq.~(\ref{type2}). It can be shown that the Ricci scalar ${\mathcal R} \sim
r^{-2}$ and the Kretschmann scalar ${\mathcal K} \sim r^{-4}$ for small $r$. There is a genuine singularity at $r=0$ which is not
covered by a horizon, i.e the singularity is naked.  According to the classification of \cite{ve2} the BST in question is a strongly naked
singularity, i.e it is not covered by a ``photon sphere'' \cite{ve3}, which is defined as a time-like hypersurface over which null geodesics which are initially
tangent to the hypersurface, remain tangents at all future times. 

We will compare geodesic motion in BSTs with that in some other relevant space-times with naked singularities.  In particular, we will be interested in the 
Janis, Newman, Winicour (JNW) \cite{janis},\cite{virbhadra} and the Joshi, Malafarina, Narayan (JMN)\cite{joshi},\cite{sahu} space-times. 
Since we deal with static, spherically symmetric space-times, we assume, without any loss of generality,
that the motion of the particle is confined to the plane $\theta = \pi/2$. It is also assumed that the background is fixed, i.e back reaction 
effects of the particle on the space-time are neglected. Further, as will be the case for all examples considered here,
we assume that $\partial_t$ and $\partial_{\phi}$ are Killing vectors, so that
\begin{equation}
\gamma = g_{tt}{\dot t}, ~~~~ h = g_{\phi\phi}{\dot \phi}
\label{conserved}
\end{equation}
are conserved quantities, corresponding to the energy and angular momentum of the particle (per unit rest mass), respectively. Here, the
dot denotes a derivative with respect to an affine parameter, which may be identified with the proper time. For time-like geodesics, we obtain
\begin{equation}
{\dot r}^2 + V\left(r\right) = 0,~~~~ V\left( r\right) =
\frac{1}{g_{rr}}\left[\frac{\gamma^2}{g_{tt}} +
  \frac{h^2}{g_{\phi\phi}} + 
1\right],
\label{potential}
\end{equation}
where $V(r)$ is an effective potential. For circular geodesics, the conditions $V(r) = V'(r)$ $=0$, where the prime denotes a derivative
with respect to  $r$, are used to solve for $\gamma$ and $h$. Stability of the circular orbit is determined from $V''(r) > 0$. For example, for the Schwarzschild solution 
with mass $M$, one obtains the effective potential and the conserved quantities
\begin{equation}
V\left(r\right) = \left(1 - \frac{2M}{r}\right)
\left(1 + \frac{h^2}{r^2}\right) - \gamma^2,~~~~\gamma^2 =
\frac{\left(r - 
2M\right)^2}{r\left(r - 3M\right)},~
h^2 = \frac{Mr^2}{r - 3M}\,.
\label{schpotential}
\end{equation}
Stability of circular orbits is then determined from the condition 
\begin{equation}
V''\left(r\right) = \frac{2M\left(6M - r\right)}{r\left(3M - r\right)} > 0\,.
\label{schstab}
\end{equation}
Eqs.~ (\ref{schpotential}) and (\ref{schstab}) show the standard textbook results that no circular orbit exists below $r=3M$ and that stable
circular orbits do not exist between $r=3M$ and $r=6M$.  Also note that from Eq.~(\ref{schpotential}), the conserved energy and momentum
diverge at $r = 3M$, the location of the photon sphere. 

Generally, solving for $\gamma$ and $h$ gives, via Eq.~(\ref{conserved}), the circular velocity
\begin{equation}
v_{\rm circ} = r \frac{d\phi}{d t} = r\frac{h}{\gamma}\frac{g_{tt}}{g_{\phi\phi}}.
\label{vcirc}
\end{equation}

A few words about the definition of the circular velocity is in order. In a GR framework, the circular velocity is a frame dependent quantity. 
An unambiguous definition of a velocity of an object in circular motion, in the context of GR can be given only in a locally flat (tetrad) basis. 
In such a basis, one projects the four momentum vector of the particle onto the tetrad frame time axis, and equates the projection
to the standard Lorentzian expression for the energy (see, e.g. \cite{hartle}). This, however, necessitates that the (stationary) 
observer is at the same radial distance as the particle undergoing geodesic motion. Although such a definition is common in the GR literature 
(see, e.g \cite{matos}, \cite{harako}), it is somewhat difficult to compare results based on this with galactic rotation curves (conventionally 
given as a function of the radial distance), as this would possibly require a series of stationary observers located at different radii. 
In our previous paper \cite{dbs1}, we proposed a Newtonian definition of the circular velocity, $v_{\rm circ} = rd\phi/dt$, but calculated via 
GR ($t$ is the co-ordinate time) and showed that this quantity gives excellent fits to galactic rotation curve data. 

Strictly speaking, $rd\phi/dt$ is not a linear velocity in the sense of GR, and our definition is phenomenological. However, $d\phi/dt$
has the usual interpretation of an angular velocity in GR, and for asymptotically flat metrics, this is the angular velocity of a particle as measured by an
observer at infinity, whose proper time coincides with the co-ordinate time \cite{hartle}. Hence, when the space-time is asymptotically flat, our prescription
of the circular velocity is closer in spirit to a GR motivated definition. With this understanding, we discuss the JMN and JNW solutions, which are
asymptotically flat but have non-zero stress tensor. 

First we consider the JMN naked singularity. In this case, the interior metric is given by
\begin{equation}
ds_{\rm JMN}^2 =  -\left(1 - M_0\right)\left(\frac{r}{r_b}
\right)^{\frac{M_0}{1 - M_0}} dt^2 + \frac{dr^2}{1-M_0} + r^2d\Omega^2\,,
\label{jmnmetric}
\end{equation}
with an external vacuum Schwarzschild solution that matches with the metric of Eq.~(\ref{jmnmetric}) at $r =r_b$. Here, $M_0$ is a parameter that
takes values between $0$ and $1$, and $d\Omega^2$ is the metric on the unit two-sphere. It can be checked that the effective potential of Eq.(\ref{potential}) is given by
\begin{equation}
V_{\rm JMN}\left(r\right) = \left(1-M_0\right)\left[1 +
  \frac{h^2}{r^2} - 
\frac{\gamma^2}{\left(1-M_0\right)}\left(\frac{r}{r_b}
\right)^{-\frac{M_0}{1 - M_0}}\right]\,.
\end{equation}
The conserved energy and angular momentum are given by 
\begin{equation}
\gamma_{\rm JMN} = \frac{\sqrt{2}\left(M_0 - 1\right)
  \left(\frac{r}{r_b}
\right)^{\frac{M_0}{2(1-M_0)}}}{\sqrt{\left(2 - 3M_0\right)}},~~~
h_{\rm JMN} = r\sqrt{\frac{M_0}{2 - 3M_0}}\,,
\end{equation}
and the angular velocity measured by an observer at infinity can be computed to be
\begin{equation}
\left(\frac{d\phi}{dt}\right)_{\rm JMN} = \sqrt{\frac{M_0}{2}} \frac{1}{r}\left(\frac{r}{r_b}
\right)^{\frac{M_0}{2\left(1 - M_0\right)}}\,.
\label{vcjmn}
\end{equation}
Note that there are no circular orbits for $M_0 > 2/3$ (the values of $M_0$ for which there are no Schwarzschild photon spheres in the
exterior region).

An entirely similar analysis follows for the asymptotically flat JNW space-time, which is sourced by a massless scalar field. Here, the metric being given by
\begin{equation}
ds^2_{\rm JNW} = - \left(1 - \frac{b}{r}\right)^ndt^2 +
\frac{1}{\left(1 - 
\frac{b}{r}\right)^n}dr^2 + r^2\left(1 - \frac{b}{r}\right)^{1-n} d\Omega^2,
\label{ds2jnw}
\end{equation}
where $0<n<1$, and there is a naked singularity at $r=b$. 
In this case, following the methods above, it is easy to check that the angular velocity is
\begin{equation}
\left(\frac{d\phi}{dt}\right)_{\rm JNW} = \frac{1}{r}\left(1 - \frac{b}{r}\right)^{n-\frac{1}{2}}\left(\frac{bn}{2r - bn - b}\right)^{\frac{1}{2}}\,.
\label{vcjnw}
\end{equation}
Similarly, for Bertrand space-times, we get \cite{dbs1}
\footnote{The BST of Eq.(\ref{type2}) used in this calculation is not asymptotically flat. However, in the next section, we will construct a solution of Einstein's equations 
with an internal BST and an external Schwarzschild metric. In that case, the angular velocity $d\phi/dt$ can be defined in GR for an observer at infinity. 
The circular velocity as per our definition of Eq.(\ref{vcirc}) will have the same qualitative form as Eq.(\ref{vcbst}).}
\begin{equation}
\left(\frac{d\phi}{dt}\right)_{\rm BST} = \frac{\sqrt{\alpha }}{\sqrt{2r}\left(\alpha + Dr\right)}\,.
 \label{vcbst}
 \end{equation}

We use the phenomenological definition in Eq.(\ref{vcirc}), to evaluate the circular velocities for the JMN, JNW and Bertrand space-times, 
using Eqs.(\ref{vcjmn}), (\ref{vcjnw}), and (\ref{vcbst}). In JMN space-times, the circular velocity increases monotonically with $r$. For JNW space-times  of Eq.(\ref{ds2jnw}),
there are three distinct ranges of $n$ for which stable circular orbits exist, for different ranges of $r$ \cite{chowdhury}, \cite{drs}. 
After a careful analysis, our main observation here is that BSTs provide the only possibility of flat rotation curves
that can match galactic data, and are thus candidates for modeling galaxies dominated by dark matter. For appropriate choices of the parameters $D$ and $\alpha$, we can 
keep the circular velocity of Eq.(\ref{vcbst}) flat, over a suitable range of $r$ \cite{dbs1}. 

In BSTs, when $r \ll \alpha/D$, the circular velocity $\sim r^{1/2}$, and matches with the special case $M_0 = 1/2$ in Eq.(\ref{vcjmn}). This also happens to be a limiting
case where the BST and the JMN metrics coincide, as the reader can verify. This indicates that BSTs can be matched with external Schwarzschild solutions. 
This is what we explicitly work out in the next section. As we have mentioned before, this will allow us to understand strong gravitational lensing in BST 
backgrounds, using the formalism of Virbhadra and Ellis. 
\footnote{The strong lensing formalism of Virbhadra and Ellis \cite{ve1}, \cite{ve2}, 
can be ideally used if the space-time is asymptotically flat, which is not the case for the metric of Eq.(\ref{type2}). 
It is certainly possible to study lensing from BSTs, in its original form in Eq.(\ref{type2}) \cite{perlickreview}, following the general formalism developed in \cite{perlickconical}. 
However,  for the purpose of connecting with experimental data, where one has
a clearly defined source and a lens, we will find it more convenient to use the theory developed in \cite{ve1}, \cite{ve2}. }

\section{Matching of BSTs with an external Schwarzschild space-time}

In this section we will discuss the matching of a BST with an external Schwarzschild space-time. The matching occurs on a time-like
hypersurface $\Sigma$, whose equation is $r=r_b$. It is well known that when two space-times are matched on a hypersurface, the first
fundamental form and the second fundamental form on $\Sigma$, computed from both the sides, must match. These facts are encoded in the
Darmois-Israel matching conditions. If the second fundamental form does not match exactly on $\Sigma$, then there emerges a thin shell of
matter around $r=r_b$ which has its own energy-momentum tensor (see, e.g \cite{poisson}). 

Specifically in this section we propose a space-time consisting of an internal BST and an external Schwarzschild solution of the form
\begin{eqnarray}
ds_{\rm BST}^2 &=& -\frac{(n+1)\beta^2}{\left(1 + \frac{nr_b}{r}\right)}c^2dt^2 + 
\frac{dr^2}{\beta^2} + r^2d\Omega^2 \,,
\label{br2n}\\
ds_{\rm SCH}^2 &=& -\left(1 - \frac{2 M_0 r_b}{r}\right) dt^2 +
\frac{dr^2}{\left(1 - \frac{2 M_0 r_b}{r}\right)} + r^2d\Omega^2
\label{scm}
\end{eqnarray}
which are matched at $r=r_b$. Here $M_0$ and $r_b$ are positive dimensionless constants, and $n$ is a positive real number. For matching
the metrics, we require the identification
\begin{eqnarray}
\beta^2 = 1 - 2M_0\,.
\label{bm} 
\end{eqnarray}
Using the Einstein equation
\begin{eqnarray}
G_{\mu \nu}=\kappa T_{\mu \nu}\,,
\nonumber
\end{eqnarray}
where conventionally $\kappa = 8\pi G_N$ and $G_N$ is Newton's gravitational constant, one can write the energy density and the
principal pressures of the matter in the interior Bertrand space-time as
\begin{eqnarray}
\rho &=&  \frac{1 - \beta^2}{\kappa r^2}\,,
\label{ed}\\
p_r &=& \frac{\beta^2(r+2nr_b)-(r+nr_b)}{\kappa r^2 (r+nr_b)}\,,
\label{pre}\\
p_\perp &=& \frac{nr_b \beta^2 (nr_b -2 r)}{4\kappa r^2 (r + n r_b)^2}\,.
\label{ppe}
\end{eqnarray}
We now demand that the weak energy condition (WEC) be satisfied in the interior BST. The energy
density is positive for $\beta < 1$, and it is convenient to write
\begin{eqnarray}
\rho + p_r &=& \frac{n\beta^2r_b}{\kappa r^2\left(r + nr_b\right)}\nonumber\\
\rho + p_{\perp} &=& \frac{4\left(1-\beta^2\right)r^2 + 2nrr_b\left(4-5\beta^2\right) + n^2r_b^2\left(4-3\beta^2\right)}{4\kappa r^2\left(r+nr_b\right)^2}\,.
\end{eqnarray}
The WEC is not satisfied for all $r$, for arbitrary values of $\beta$, as can be seen by taking $\beta \to 1^-$. To avoid this 
difficulty, we will choose $\beta^2 < 4/5$, so that the WEC is satisfied for all values of the radial distance. 

Calculating the extrinsic curvature on both sides of the junction, it can be shown that
these match at the hypersurface $r=r_b$ \footnote{Equivalently, the junction condition can be stated as the requirement $p_r=0$ at $r=r_b$.} for 
\begin{equation}
\beta^2 = \frac{n+1}{2n+1},~~~~\implies M_0 = \frac{n}{2\left(1+2n\right)}
\label{di}
\end{equation}
where the second identity is obtained from Eq.(\ref{bm}), and the WEC dictates that $n > 1/3$. 
If Eq.(\ref{di}) is not satisfied then although the first fundamental form matches at the boundary, the second 
fundamental form does not, and this implies a thin shell of matter at the boundary. \footnote{Note that what ever be the matching conditions, Eq.~(\ref{bm}) always holds.}
In this case, a surface stress-energy tensor develops on both sides of $\Sigma$, given by \cite{poisson}
\begin{eqnarray}
S_{ab}=-\frac{1}{8\pi}\left([K_{ab}] - [K] h_{ab}\right)\,,
\label{sab}
\end{eqnarray}
where $a$, $b$ are space-time indices on $\Sigma$ and $h_{ab}$ is the induced metric on the hypersurface. Here, $K$ is the
trace of the extrinsic curvature, and $[K]=K^{\rm SCH}|_\Sigma - K^{\rm BST}|_\Sigma$, where the superscripts
indicate the region in which $K$ is to be calculated.  As we will not
use the explicit form of the stress-energy tensor in this paper we do not go into further details of the form of $S_{ab}$. 

The Schwarzschild radius in our solution is $r_s = 2M_0r_b$. If we match the two solutions of Eqs.(\ref{br2n}) and (\ref{scm}) without a thin shell at the
boundary, then Eqs.(\ref{scm}) and (\ref{di}) imply that 
\begin{equation}
r_s = \frac{nr_b}{1+2n}\,.
\end{equation}
Hence, the Schwarzschild radius stays inside the matching radius for all values of $n$, and asymptotes to $r_b/2$ as $n$ is made large.
On the other hand, if there is a thin shell of matter at the boundary, then for small values of $\beta$, $M_0 \to 1/2$, and the
Schwarzschild radius, in this limit, approaches $r_b$ from inside. However, the radius of the photon sphere, as we will see in the next 
section, can be greater than $r_b$. 

To make contact with experimental data, we demand that our metric of Eqs.(\ref{br2n}) and (\ref{scm}) give rise
to realistic values for the galactic rotation curves and the galactic mass. As we have stated, the WEC implies the 
condition $n > 1/3$ and we will now make a specific choice, $n=2$. This will not change any qualitative aspects 
of our results, while simplifying the discussion considerably. In this case, $\beta^2 = 3/5$ and 
consequently, $M_0 = 1/5$. Both $p_r$ and $p_{\perp}$ vanish at the boundary, and there is no mismatch of pressure on the matching hypersurface. 
If these values of $\beta$ and $M_0$ are not chosen, then we have a thin shell of matter at the matching surface. 

Now, let us discuss the rotation curves in our interior BST, as seen by an external Schwarzschild observer at infinity. With our choice of $n=2$, the circular velocity 
for time-like geodesics, with our definition of the last section, can be shown to be given by (restoring the speed of light, $c$ in the metric of Eq.(\ref{br2n}))
\begin{eqnarray}
v_{\rm circ} =  \left.\frac{c \beta \sqrt{n(n+1) r r_b}}{\sqrt{2}(r+nr_b)}\right|_{n=2} = \frac{c \beta \sqrt{3 r r_b}}{r+2r_b}
\label{vcirc1}
\end{eqnarray}
where the first expression is the circular velocity is calculated from Eq.(\ref{br2n}).
\begin{figure}[t!]
\centering
\includegraphics[width=8cm,height=6cm]{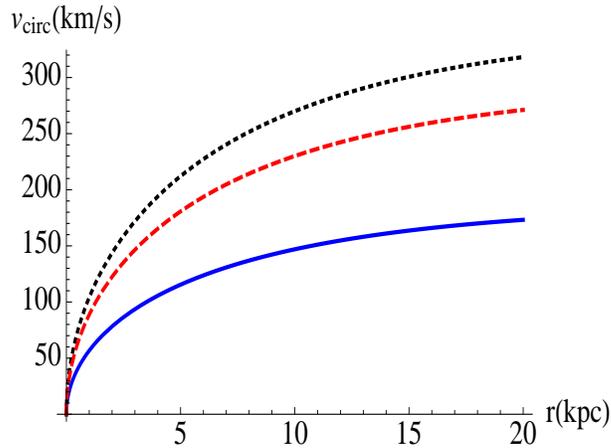}
\caption{Color Online : Figure showing the velocity rotation curves for small $\beta$. For details, see text.}
\label{vcn}
\end{figure}
Thus, if we demand that $v_{\rm circ}\sim 10^2\,{\rm km/s}$ at $r \sim r_b$, then we should take $\beta \sim 10^{-3}$. This value of $\beta$ does not satisfy the
Darmois-Israel junction conditions but will give us interesting predictions about gravitational lensing, as we will see in the next section. 
In general, $v_{\rm circ}$ maximizes at $r = nr_b$. For $n=2$, this would imply that the rotation curves are monotonically rising inside
the galactic region. Such rising galactic rotation curves are known in the literature in the context of LSB galaxies \cite{rising1},\cite{rising2} - for example
the galaxy NGC4594, popularly called the ``sombrero'' galaxy - and our model fits well with these cases. \footnote{For $n < 1$, the flat region of the galactic rotation 
curve falls within the BST. Hence in our model, matching the galactic rotation curves will require a choice of $n$. In this paper, we only discuss the case $n=2$.} 
Typically, the slope of the galactic rotation curve is high for galaxies with large $r_b$, as shown in Fig.(\ref{vcn}). Here, for the solid blue curve, 
$\beta=10^{-3}$ and $r_b=20~{\rm kpc}$. For the dashed red curve, $\beta=1.4\times 10^{-3}$ and $r_b=25\,{\rm kpc}$. Finally, for the dotted black curve, 
$\beta=1.5\times10^{-3}$ and $r_b=30\,{\rm kpc}$. The fact that bigger galaxies have steeper rotational velocity curves, fits well with experimental data. 

Let us now turn to the range of galactic mass in our model. The Schwarzschild mass of the galaxy, in our model, is given by  $M=M_0r_b$. If we demand
that the matching condition of our solution does not have a thin shell at the boundary then $M_0 = 1/5$ and $M=r_b/5$. 
If we choose a typical matching radius $r_b \sim 20~{\rm kpc}$ and then restore normal units, \footnote{We will work in standard geometric units, in which the
Newton's constant is set to $G = 4.3 \times 10^{-3} ({\rm pc}/M_{\odot}) {\rm (Km/sec)}^2$.}  we obtain $M \sim 10^{17}M_{\odot}$, i.e the Schwarzschild 
mass of the interior region turns out to be four to five orders greater than the general mass range of heavy galaxies. If we assume matching with a
thin shell at the boundary and take $\beta = 10^{-3}$, i.e $M_0 \sim 1/2$, the Schwarzschild mass does not change appreciably. 

It is instructive to look at the Newtonian density profile with
the circular velocity that we discussed above. This is given by (the subscript $N$ indicates that we are working in the Newtonian approximation)
\begin{equation}
\rho_N\left(r\right) = \frac{1}{4\pi G_Nr^2}\frac{d}{dr}\left(v_{\rm circ}^2r\right) = \frac{c^2\beta^2r_b^2 n\left(n+1\right)}{2G_N\pi r\left(r + 2r_b\right)^3}\,.
\label{den}
\end{equation}
This is a Hernquist density profile, but reduces to the NFW profile near the maximum of $v_{\rm circ}$, as discussed in \cite{dbs1}. Integrating
this profile up to $r=r_b$, the Newtonian mass of the galaxy is given by
\begin{equation}
M_N  = \frac{\beta^2c^2n\left(n+1\right)r_b}{18G_N}\,.
\end{equation}
Now, with $\beta = \sqrt{3/5}$ and $n=2$, we obtain $M_N \sim 10^{17}M_{\odot}$ for $r_b = {\rm 20Kpc}$. However, when $\beta \sim 10^{-3}$, we obtain 
$M_N \sim 10^{11}M_{\odot}$, which is closer to observed galactic masses. The Schwarzschild mass as seen by an external observer at infinity, is however 
not sensitive to such changes to $\beta$, as alluded to before. The difference between this and the $M_N$ can thus be interpreted as contributing to the 
gravitational binding energy and the energy due to matter in the thin shell. 

Importantly, in order to get a standard Schwarzschild mass range of $\sim 10^{12}M_{\odot}$ with a typical matching radius of around $20 {\rm kpc}$, 
we need to choose a small value of $M_0$, i.e of $n$.  From Eq.(\ref{di}), this will however imply a value of $\beta$ close to unity, which in turn violates the WEC, in
certain ranges of $r$. On the other hand, since $M = M_0r_b$, a possible way to get a reasonable mass estimate is to choose a small value of $r_b$. 
In that case $r_b$ should not be thought of as the actual galactic radius, but an effective radius within which the effect of the dark matter and its associated 
naked singularity is maximized, and outside this one can approximate the metric as effectively Schwarzschild. 

Having constructed a solution that is internally a BST and externally Schwarzschild, we are now ready to study strong gravitational lensing in this space-time. 
This is what we undertake in the next section. 

\section{Gravitational Lensing in Bertrand space-times}

In this section, we study gravitational lensing from BSTs (for an excellent review, see \cite{perlickreview}), and in particular, we focus on  
Einstein rings (ERs). We use the strong gravitational lensing formalism of \cite{ve1}, \cite{ve2} and follow their notation. 
For a static, spherically symmetric metric of the form
\begin{equation}
ds^2 = -A(r)dt^2 + B(r)dr^2 + r^2d\Omega^2
\end{equation}
where $d\Omega^2$ is the standard metric on the two sphere, it can be shown that the Einstein deflection angle of light is given as 
\begin{equation}
\alpha_D(r_0) = 2\int_{r_0}^{\infty} \left[B(r)\right]^{\frac{1}{2}}\left[\left(\frac{r}{r_0}\right)^2\frac{A(r_0)}{A(r)} -1 \right]^{-\frac{1}{2}}\frac{dr}{r} - \pi\,.
\label{bend}
\end{equation}
In terms of $r_0$, which is the closest distance of approach of the light ray, the impact parameter, denoted by $J$, is given by
\begin{equation}
J = r_0\left[A(r_0)\right]^{-\frac{1}{2}}\,.
\label{impact}
\end{equation}
\begin{figure}[t!]
\centering
\includegraphics[width=2.5in,height=3.2in]{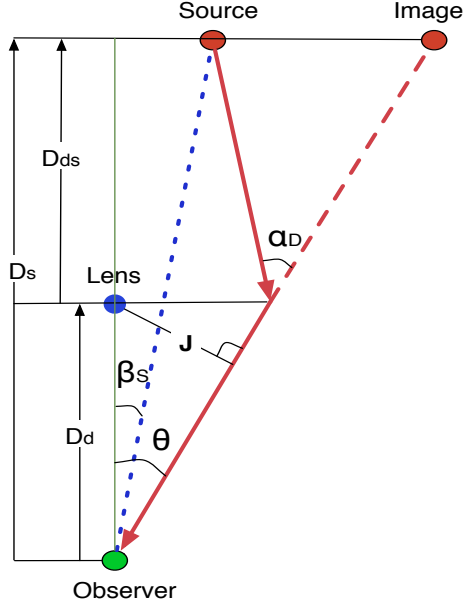}
\caption{Color Online : Diagram for strong gravitational lensing, reproduced from \cite{ve1}. Details are given in the text.}
\label{lensf}
\end{figure}
These have to be used in conjunction with the Virbhadra-Ellis lens formula for strong gravitational lensing, given by
\begin{equation}
{\rm tan} \beta_s = {\rm tan} \theta - {\tilde\alpha} ;~~~~{\tilde\alpha} = \frac{D_{ds}}{D_s}\left[{\rm tan}\theta + {\rm tan}(\alpha_D - \theta)\right]\,,
\label{lens}
\end{equation}
where the terms in the above equations are explained in Fig.~\ref{lensf}, which has been reproduced from \cite{ve1}. The solid red lines indicate light rays
that undergo strong gravitational lensing from the source. The image position is depicted by the dashed red line. The dotted blue line is the original position of 
the source, and to consider ERs, we need to set $\beta_s = 0$. The distance between the lens and the 
observer is $D_{d}$, between the source and lens is $D_{ds}$ and between the source and the observer is
$D_s$. The impact parameter for light is $J$. Angles are measured from the vertical line joining the lens L and observer O. The source makes
an angle $\beta_s$ and the image makes an angle $\theta$ with the vertical. In the diagram, $\alpha_D$ is the angle of deflection in the
thin lens approximation. From the lens diagram, one gets $J = D_d \sin\theta$. From this, and Eqs. (\ref{bend}) and (\ref{impact}), one can obtain
the bending angle $\alpha_D$ as a  function of $\theta$. For completeness, let us also state that the magnification of images is given by the formula
$\mu = \mu_r\mu_T$, where $\mu_r$ and $\mu_T$ are the radial and
tangential magnifications, given by
\begin{equation}
\mu_r = \frac{d\theta}{d\beta_s};~~~~~\mu_T = \frac{{\rm sin}\theta}{{\rm sin}\beta_s}\,.
\label{mageq}
\end{equation}

The above formalism can be used to compute image locations and their magnifications, for strong gravitational lensing. For example, if we plot 
$\left({\rm tan}\theta - {\rm tan}\beta_s\right)$ and ${\tilde\alpha}$ as functions of $\theta$ for various fixed values of $\beta_s$, their intersection gives us the image positions for those 
values of $\beta_s$. As mentioned, when $\beta_s=0$, i.e the observer is aligned with the lens and the source, the resulting cylindrical symmetry dictates that an Einstein
ring is formed. Importantly, one distinguishes between relativistic and non-relativistic images and ERs. Relativistic images are said to form if the bending angle of
light is more than $3\pi/2$ and similarly, relativistic ERs refer to rings for which the bending angle is greater than $2\pi$. It might seem
that relativistic ERs can only form when the lens is a black hole, i.e when light goes around the photon sphere a number of times before reaching
the observer, however, it has been shown \cite{sahu} that such large bending angle of light can be obtained for strongly naked singularities,
which do not have a photon sphere. 

Since an important role in our analysis will be played by the photon sphere, let us discuss this in some details. From what we have said in section 2, A photon
sphere is a time-like hypersurface where the bending angle of light as a function of the distance of closest approach
diverges, as this distance approaches the radius of the sphere. In a pure BST, as mentioned in \cite{dbs1}, the photon sphere does not
exist, and hence this is an example of a strongly naked singularity, according to the classification of \cite{ve2}. In the context of our solution, 
given in Eqs. (\ref{br2n}), (\ref{scm}), a more detailed analysis is required,
since the external Schwarzschild solution supports a photon sphere. Specifically, for the Schwarzschild solution in our case
(where the metric is as given in Eq.~(\ref{scm})), the event horizon is at $r_s = 2M_0 r_b$ and the photon sphere associated with this
Schwarzschild space-time is at $r_{\rm ps}=3M_0 r_b$. The importance of the photon sphere in this case lies in the fact that the Einstein
bending angle of light is finite only when the distance of closest approach is greater than the radius of the photon sphere.  Introducing
a dimensionless variable $x = r/(2M) = r/(2M_0 r_b)$ (which is extensively used in sequel), the location of the photon sphere is equivalently at
$x_{\rm ps} = 3/2$, whereas our BST extends up to the matching radius $r = r_b$, i.e $x_b = r_b/(2M_0 r_b) = 1/(2M_0)$.  

From the discussion above it is seen that in our case we have two distinct length scales, i.e the matching radius $r_b$ (or $x_b$), and the 
radius of the photon sphere $r_{\rm ps}$ (or $x_{\rm ps}$), which play important roles in gravitational lensing (the Schwarzschild radius is always within the matching 
radius, as we have said). Depending on their magnitudes, two physically interesting situations might arise, 
which are summarized as : \footnote{Note that we have set $n=2$ in Eqs.(\ref{br2n}) and (\ref{scm}). Even if this is not done, we get the same
conditions as presented here.}
\begin{enumerate}
\item 
The photon sphere (of the external Schwarzschild space-time) may be inside the matching radius (so that effectively there is no photon sphere). 
This happens when $x_b > x_{\rm ps}$, which implies 
\begin{equation}
\frac{1}{2M_0} > \frac{3}{2}  \implies M_0 < \frac{1}{3} 
\implies \beta > \frac{1}{\sqrt{3}}\,,
\label{condns}
\end{equation}
where the relation $\beta^2 = 1-2M_0$ has been used. In this case light from the distant source enters the BST region 
and then deviates its path, due to the lens, and reaches the observer. This situation arises when we match the metrics of Eqs. (\ref{br2n}) and (\ref{scm}) 
with the Darmois-Israel junction conditions.  
\item 
If $M_0 > \frac{1}{3}$ (or $\beta < \frac{1}{\sqrt{3}}$), then the photon sphere of the Schwarzschild space-time goes outside the region covered by BST. 
In this case light coming from the distant source never reaches the inner BST region, gets deflected outside the photon sphere of the external Schwarzschild region, 
and then reaches the observer. This condition may arise when the matching of the internal BST and the external Schwarzschild solution is through a thin 
shell around $r=r_b$, i.e the junction conditions are not satisfied. 
\end{enumerate}
\begin{figure}[t!]
\begin{minipage}[b]{0.5\linewidth}
\centering
\includegraphics[width=2.8in,height=2.8in]{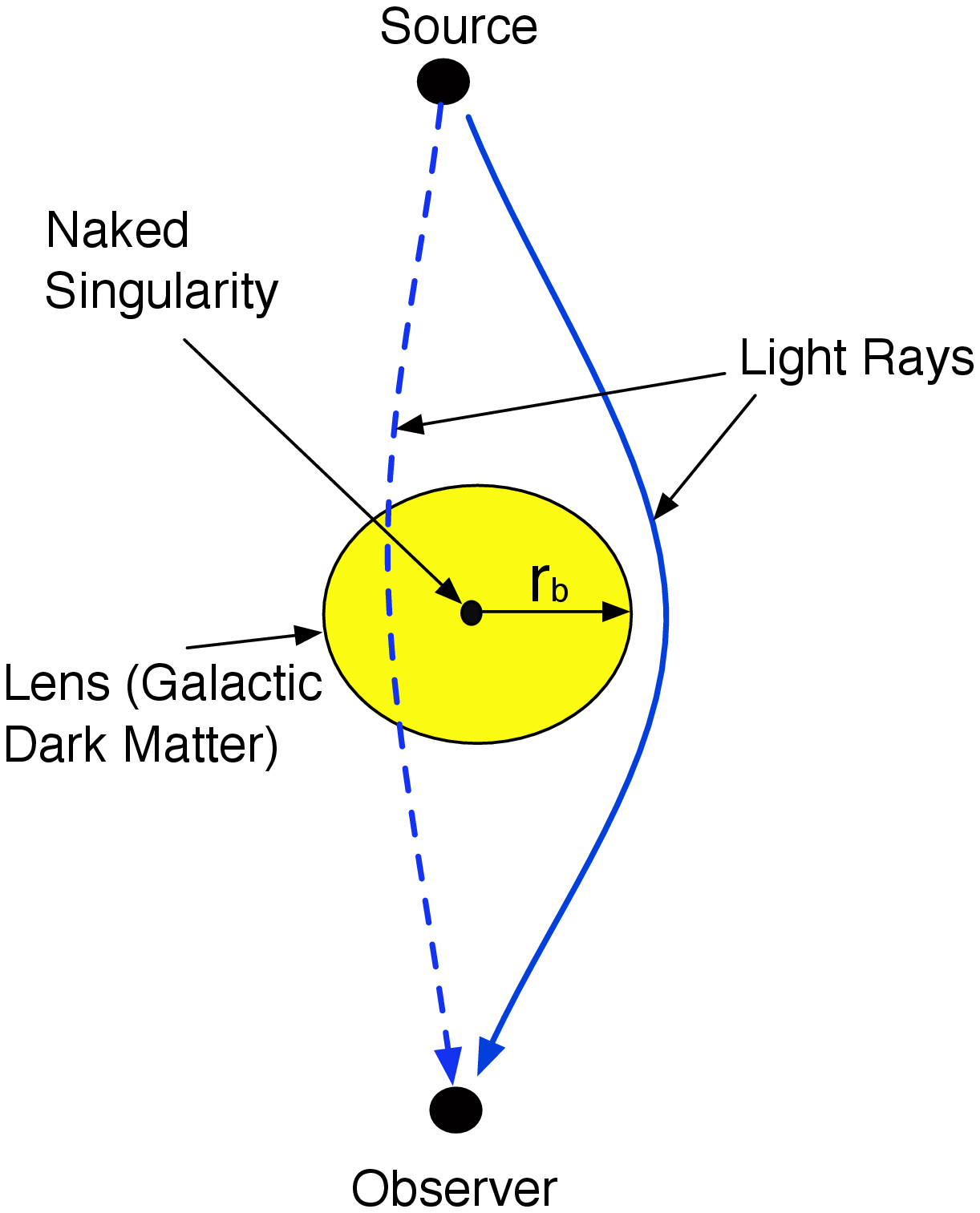}
\caption{Color Online: When there is no external photon sphere, some light rays (dashed blue line) from the source can reach the observer
passing through the dark matter dominated region.}
\label{qual1}
\end{minipage}
\hspace{0.2cm}
\begin{minipage}[b]{0.5\linewidth}
\centering
\includegraphics[width=2.8in,height=2.8in]{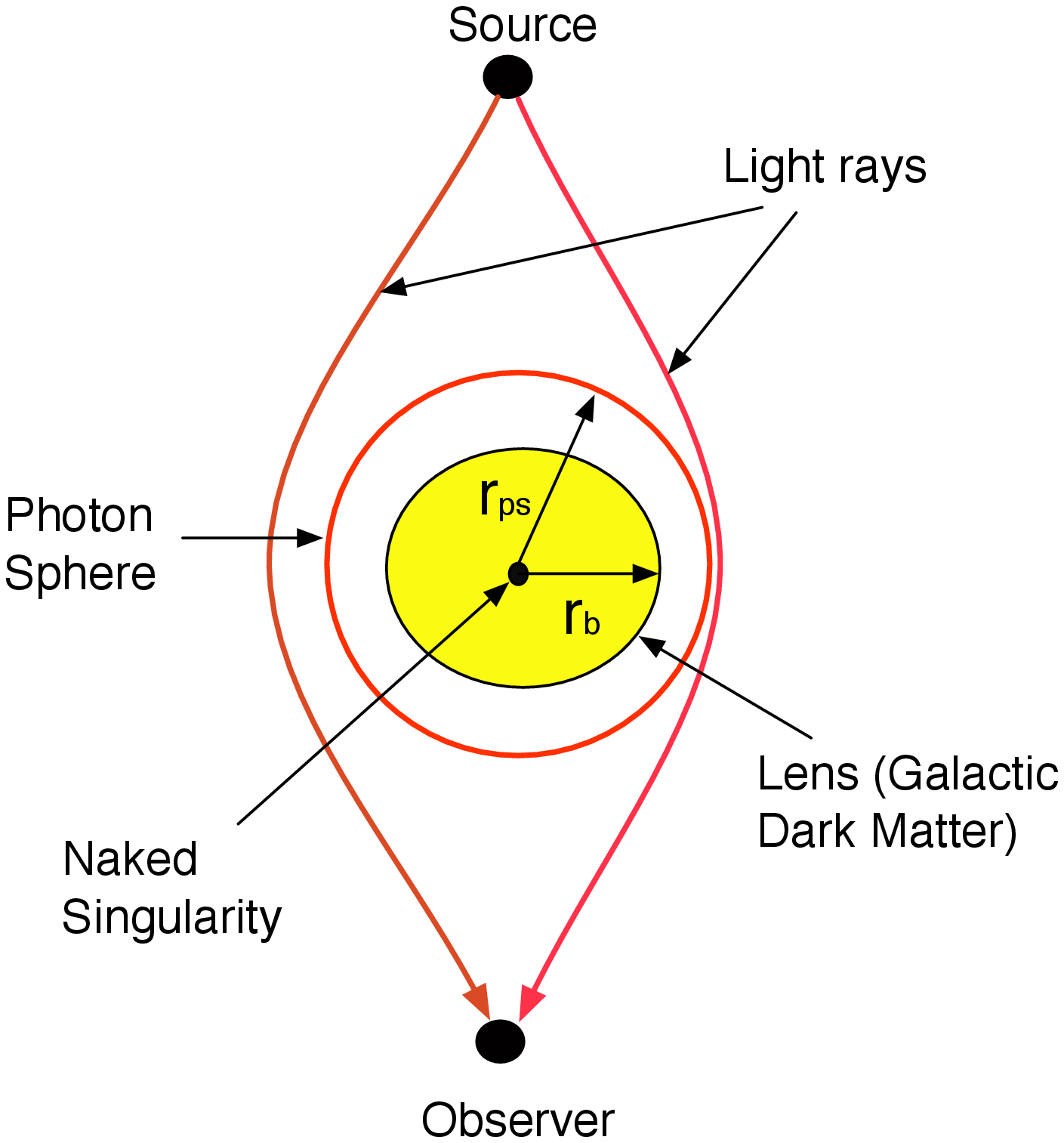}
\caption{Color Online : In the presence of a photon sphere outside the matching radius, light rays reach the observer without entering the
interior dark matter region.}
\label{qual2}
\end{minipage}
\end{figure}
A qualitative depiction of these two situations is presented in Figs.(\ref{qual1}) and (\ref{qual2}). In Fig.(\ref{qual1}), there is no photon sphere, and
light rays can come arbitrarily close to the central naked singularity, as depicted by the dashed blue line. In Fig.(\ref{qual2}), due to the external photon
sphere, this is not possible, and as far as gravitational lensing is concerned, the photon sphere shields the effect of the internal space-time. 

We will now discuss these two cases quantitatively. The mathematical formalism involves Eqs.(\ref{bend}) and (\ref{impact}), in conjunction with 
Eqs. (\ref{br2n}) and (\ref{scm}) (and the fact that $J = D_d{\rm Sin}\theta$), where we set $n=2$, to make the discussion simple. If the light ray enters the 
internal BST, then the deflection angle is given by 
\begin{eqnarray}
\alpha_D &=& 2\int_{x_0}^{\frac{1}{2M_0}}{\mathcal A}_{\rm BST} dx + 2\int_{\frac{1}{2M_0}}^{\infty}{\mathcal B}_{\rm SCH} dx - \pi \nonumber\\
{\mathcal A}_{\rm BST} &=& \frac{1}{\beta x}\left[\frac{(x-x_0)\left(1+
(x+x_0)M_0\right)}{x_0(1+M_0x_0}\right]^{-\frac{1}{2}}\nonumber\\
{\mathcal B}_{\rm SCH} &=& \frac{1}{x}\left[\frac{(3\beta^2M_0x^3 - x_0(x-1)(1+M_0x_0))}{x x_0(1+M_0x_0)}\right]^{-\frac{1}{2}}
\label{lens1}
\end{eqnarray}
along with the image position 
\begin{equation}
{\rm sin}\theta = \frac{2Mx_0}{\beta\sqrt{3}D_d}\left(1 + \frac{1}{x_0 M_0}\right)^{\frac{1}{2}}
\label{lens2}
\end{equation}
Here, ${\mathcal A}_{\rm BST}$ and ${\mathcal B}_{\rm SCH}$ are the deflection angles in the internal BST and the external Schwarzschild region, respectively,
and $x_0$ is taken to be small compared to $x_b$.
The internal region will not contribute if the photon sphere due to the external Schwarzschild solution lies outside the matching radius, as we have mentioned. 
In this case, the deflection angle, which is purely due to the external Schwarzschild solution, can be calculated from the analysis of \cite{ve1}. 
\begin{figure}[t!]
\begin{minipage}[b]{0.5\linewidth}
\centering
\includegraphics[width=2.8in,height=2.3in]{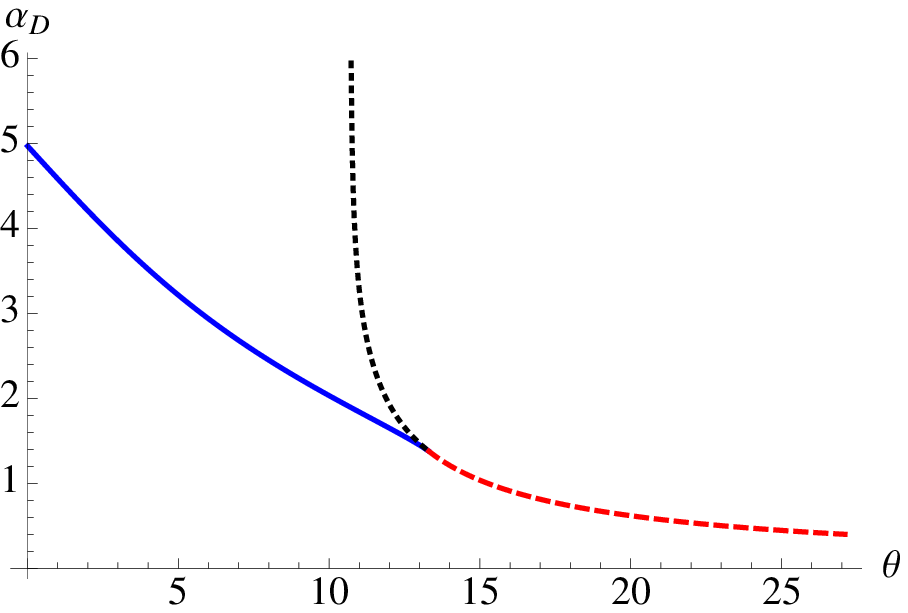}
\caption{Color Online : Angle of deflection of light as a function of the image position $\theta$ for the galaxy cluster Abell 370. $\alpha_D$ is in radians
and $\theta$ is in milli arcseconds.}
\label{angle1}
\end{minipage}
\hspace{0.2cm}
\begin{minipage}[b]{0.5\linewidth}
\centering
\includegraphics[width=2.8in,height=2.3in]{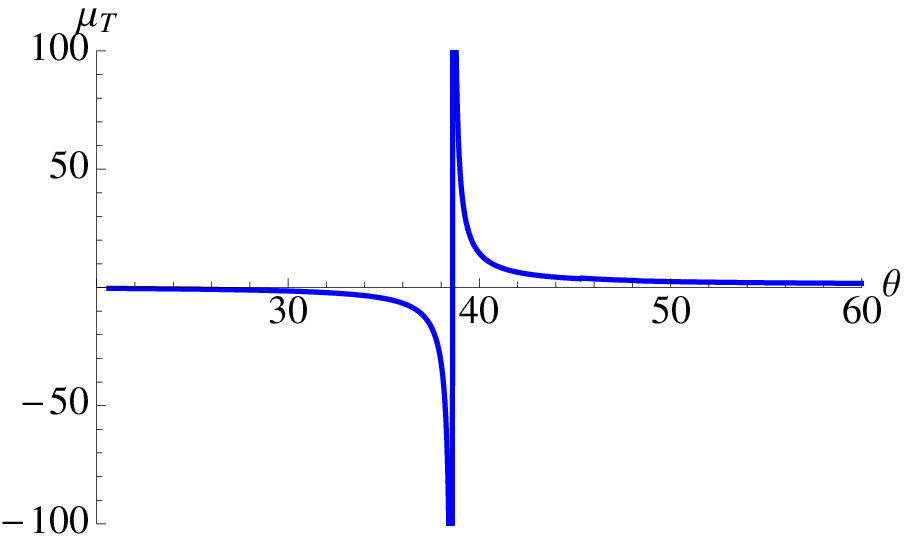}
\caption{Color Online : Tangential magnification as a function of $\theta$, for the galaxy cluster Abell 370. Here, $\theta$ is measured in arcseconds.
$\mu_T$ diverges at the location of the ER.}
\label{mag1}
\end{minipage}
\end{figure}
Having obtained the deflection angle, the location of Einstein rings can be obtained by graphically solving Eq.(\ref{lens}), after setting $\beta_s = 0$,
or equivalently, calculating the locations of the divergences of $\mu_T$ in the second relation of Eq.(\ref{mageq}). 

We first analyze gravitational lensing in case (1). Here, we set set $M_0 = 1/5$, so that there is no matter at the matching
hypersurface. The Schwarzschild mass is taken as $\sim 10^{15}M_{\odot}$, and assuming a typical distance of $10^6 {\rm Kpc}$ to the lens, we get 
$M/D_d \sim 10^{-8}$. We have chosen these values, keeping in mind the lensing data for the cluster Abell 370 \cite{abell}, 
\footnote{Although most of the results in this paper focus on galactic dark matter, one can also work out the strong lensing 
behavior from a galaxy cluster, as Abell 370 \cite{abell} (and probably others), with our techniques.} where the source galaxy
is at a redshift of $z=2$, and the lens is at a redshift $z=0.375$. Since the Schwarzschild mass of the cluster 
is $10^{15}M_\odot $, we get $r_b \sim 0.5\,\,{\rm kpc}$. \footnote{If we take the dark matter mass to be $\sim 10^{17}M_{\odot}$, then, with
$M_0 = 1/5$, the matching radius is $\sim 20~{\rm kpc}$. In this case, the radius of the ER is $\sim 650~{\rm ArcSec}$. Such a giant ring has not been
observed till now.} 
This implies that the effect of the central singularity essentially spreads only up to $0.5\,{\rm kpc}$. While applying our model to galaxy clusters, it should be assumed that 
most of the galaxies in the cluster are themselves effectively in the outer Schwarzschild region, and in the interior region of the cluster there is a dark matter 
dominated region hiding the naked singularity. 

In Fig.~(\ref{angle1}), we show the deflection angle (in radians) as a function of the image position (in milli ArcSec). The solid blue line represents the deflection angle when
light rays enter the BST region. The continuation of this with increasing $\theta$, as the dashed red line is for the case where the bending takes place entirely in the
Schwarzschild region. \footnote{The continuation of the dashed red Schwarzschild line with decreasing $\theta$ as the dotted black line, represents the situation with no interior
BST (i.e only a Schwarzschild solution with mass $10^{15}M_{\odot}$), in which case the deflection angle would diverge at the Schwarzschild photon sphere. }
In our case, the photon sphere lies inside the BST. We see that here, relativistic ERs are not possible. The only ER would be due to the bending of light in the external region. In 
Fig.~(\ref{mag1}), we have shown the tangential magnification $\mu_T$, obtained from Eq.(\ref{mageq}), by choosing $D_{ds}/D_s =0.875$ \cite{abell}. 
This blows up at $\sim 39''$, indicating the ER. We note here that giant rings have been observed for the galaxy cluster Abell 370 at almost the same value 
of the radius \cite{abell}. Also, in this case, by choosing $\beta = 10^{-3}$, it can be checked that the position of the ER does not change. This is because, with this choice 
of $\beta$, the radius of the photon sphere is outside the BST region, and the lensing is purely due to the external Schwarzschild space-time. 
\begin{figure}[t!]
\begin{minipage}[b]{0.5\linewidth}
\centering
\includegraphics[width=2.8in,height=2.3in]{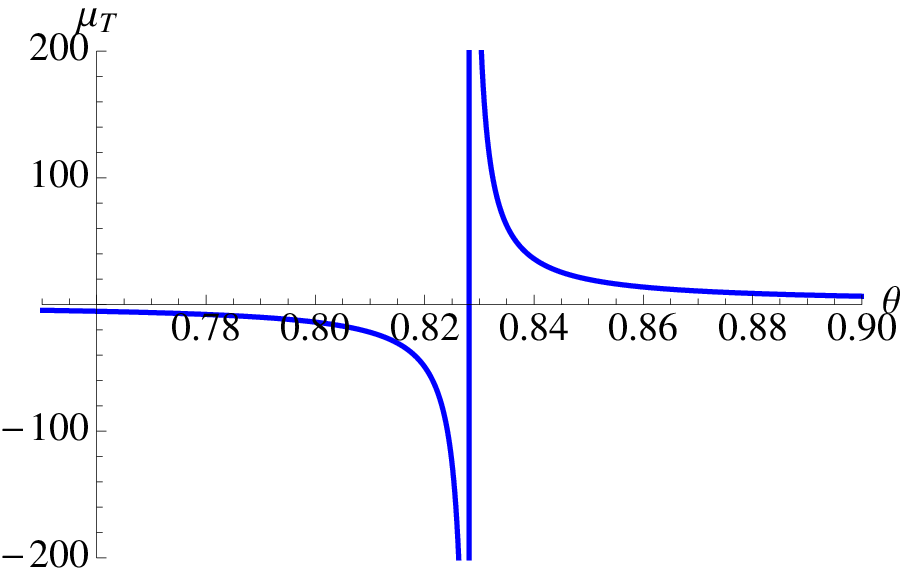}
\caption{Color Online : Tangential magnification as a function of $\theta$ (in arcseconds), for the lens system MG 1654 + 134.}
\label{mag2}
\end{minipage}
\hspace{0.2cm}
\begin{minipage}[b]{0.5\linewidth}
\centering
\includegraphics[width=2.8in,height=2.3in]{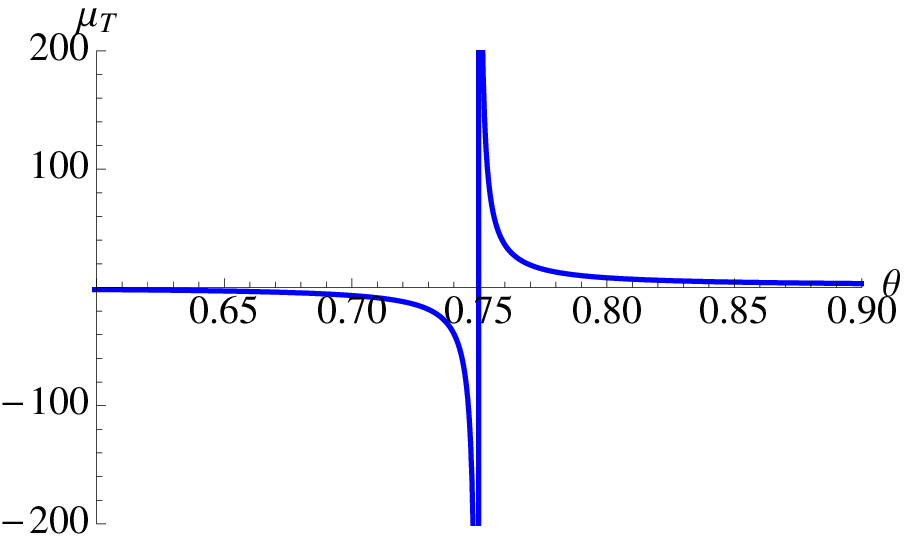}
\caption{Color Online : Tangential magnification as a function of $\theta$ (in arcseconds), for the lens system MG 1131 + 0456.}
\label{mag3}
\end{minipage}
\end{figure}

Finally, for case (2), we have considered two known lens systems, MG 1654 + 134 \cite{kochanek}, and MG 1131 + 0456 \cite{hewitt}. For these two
systems, $M/D_d$ is taken as $\sim 4.8 \times 10^{-12}$ and $\sim 1.32 \times 10^{-11}$ respectively. The mass estimates of the galaxies are known to be
be $\sim 10^{11} M_{\odot}$ and $\sim 10^{12} M_{\odot}$ respectively. In Figs.~(\ref{mag2}) and (\ref{mag3}), we have shown the location of the 
ER for this data, where we have chosen $\beta = 10^{-3}$, so that the lensing is effectively from the external Schwarzschild region. 
We find that for the system MG 1654 + 134, the predicted value of the ER is at $\theta \sim 0.83''$, while it is at $\theta \sim 0.75''$ for MG 1131 + 0456.
This is close to observed values of $1''$ and $0.87''$ for these galaxies, respectively \cite{kochanek},\cite{hewitt}. These results indicates that as
far as gravitational lensing is concerned, galactic dark matter may behave similarly as Schwarzschild backgrounds because of the presence of
the photon sphere. 

\section{Discussions and Conclusions}

In this paper, we have attempted to construct a viable GR model for galactic dark matter. 
We first compared BSTs with other space-time singularities, and established the fact that as far as galactic rotation curves are concerned, there is strong
reason to model dark matter dominated galactic space-time with BSTs. We then studied gravitation lensing in BSTs. In particular, we discussed strong 
gravitational lensing, following the formalism of Virbhadra and Ellis \cite{ve1}, \cite{ve2}. The results seem to be in good agreement with observational data. 
Our main conclusion in this paper is that realistic modeling of galaxies is possible using BSTs. 

We also calculated the lensing properties of a galaxy cluster, such as Abell 370, and observed that although BSTs are mainly thought of as galactic models, 
but for lensing properties, the BST with an external Schwarzschild solution can predict characteristics from galaxy clusters as well. Also, our analysis reveals that as 
far as lensing is concerned, the effect of the central naked singularity in the BST may be shielded by a photon sphere of the external Schwarzschild solution, as seen by an 
observer at asymptotic infinity. It is tempting to think that this might provide a reason as to why a vacuum Schwarzschild solution provides good estimates for observed 
values of the radii of Einstein rings, in different galactic lens systems. 

As we have mentioned before, one can study lensing directly in a BST (i.e without a matching asymptotically flat solution) by using the formalism of 
Perlick \cite{perlickconical}. It will be very interesting to see if this can be used in the context of galactic astrophysics. 

In conclusion, we have shown that Bertrand space-times can have interesting astrophysical applications. Alone, it can serve as a model of dark matter dominated,
low surface brightness galaxies. In conjunction with an external Schwarzschild solution, it can be used to calculate strong lensing effects from such galaxies. This 
setup can also loosely mimic lensing from a galaxy cluster. It might be possible to make a more realistic galactic model with a rotating Bertrand space-time. 
We leave this for a future publication.

\begin{center}
{\bf Acknowledgements}
\end{center}
It is a pleasure to thank Sayan Kar for extremely useful comments.



\begin{thebibliography}{999}
\bibitem{nfwprof} 
J.~F.~Navarro, C.~S.~Frenk, S.~D.~M.~White, Astrophys.\ J.\  {\bf 462}, 563 (1996)
[astro-ph/9508025].

\bibitem{sayan} 
S.~Bharadwaj, S.~Kar, Phys.\ Rev.\ D {\bf 68}, 023516 (2003).

\bibitem{mdroberts}
M. D. Roberts, Gen. Rel. Grav. {\bf 36} no. 11, 2423 (2004).

\bibitem{dbs1}
D.~Dey, K.~Bhattacharya and T.~Sarkar, Phys. Rev. {\bf D87} 103515 (2013).

\bibitem{binneytremaine}
J. Binney, S. Tremaine, ``Galactic Dynamics,'' Princeton University Press, 2008.

\bibitem{perlick} 
V.~Perlick, Class.~Quantum~Grav., {\bf 9} (1992) 1009.

\bibitem{bert}
J. Bertrand, Compt. Rend. {\bf 77} (1873) 849.

\bibitem{goldstein} H.~Goldstein, ``Classical Mechanics'', $2^{\rm nd}$ edition,
Narosa Publishing House (1993).

\bibitem{kb}
P.~Kumar, K.~Bhattacharya, Eur.\ J.\ Phys.\  {\bf 32}, 895 (2011).

\bibitem{kbs1} 
P.~Kumar, K.~Bhattacharya, T.~Sarkar, 
Phys.\ Rev.\ D {\bf 86}, 044028 (2012).

\bibitem{kbs2} P.~Kumar, K.~Bhattacharya, T.~Sarkar, ``Geodesic flows and their deformations in Bertrand space-times,'' 
{\tt arXiv:1208.5327 [gr-qc]}.  (Talk given at the 13th Marcel Grossmann Meeting on Recent Developments in Theoretical and
Experimental General Relativity, Astrophysics, and Relativistic Field Theories, Stockholm, 2012).

\bibitem{salucci}
P. Salucci, A. Lapi, C. Tonini, G. Gentile, I. Yegorova, U. Klein, MNRAS, 378 41 (2007). 

\bibitem{hernquist}
L. Hernquist, Astrophys. J {\bf 356}, 359 (1990).

\bibitem{ve1}
K.~S.~Virbhadra, G.~F.~R.~Ellis, Phys.\ Rev.\ D {\bf 62}, 084003 (2000).

\bibitem{ve2} 
K.~S.~Virbhadra, G.~F.~R.~Ellis, Phys.\ Rev.\ D {\bf 65}, 103004 (2002).

\bibitem{v1}
K. S. Virbhadra, D. Narashimha, S. M. Chitre, Astron. Astrophys. {\bf 337}, 1 (1998).

\bibitem{v2}
K. S. Virbhadra, C. R. Keeton, Phys. Rev. {\bf D77}, 124014 (2008). 

\bibitem{v3}
K.~S.~Virbhadra, Phys. Rev. {\bf D79}, 083004 (2009).

\bibitem{bozza1}
V.~Bozza and L.~Mancini, Astrophys.\ J.\  {\bf 696}, 701 (2009).

\bibitem{bozza2}
V.~Bozza and L.~Mancini, Astrophys.\ J.\  {\bf 611}, 1045 (2004).

\bibitem{ve3}
C.~-M.~Claudel, K.~S.~Virbhadra, G.~F.~R.~Ellis,  J.\ Math.\ Phys.\  {\bf 42}, 818 (2001)

\bibitem{janis} 
A.~I.~Janis, E.~T.~Newman, J.~Winicour, Phys.\ Rev.\ Lett.\  {\bf 20}, 878 (1968).\\
M. Wyman, Phys. Rev. D 24, 839 (1981).\\

\bibitem{virbhadra}
K. S. Virbhadra, Int J Mod Phys {\bf A12} 4831 (1997).

\bibitem{joshi}
P.~S.~Joshi, D.~Malafarina, R.~Narayan,
Class.\ Quant.\ Grav.\  {\bf 28}, 235018 (2011).

\bibitem{sahu} 
S.~Sahu, M.~Patil, D.~Narasimha, P.~S.~Joshi, 
Phys.\ Rev.\ D {\bf 86}, 063010 (2012).

\bibitem{hartle}
J. B. Hartle, ``Gravity - An Introduction to Einstein's General Relativity,'' Pearson Education Inc. (2003).

\bibitem{chowdhury}
A. N. Chowdhury, M. Patil, D. Malafarina, P. S. Joshi, Phys. Rev. D 85, 104031 (2012).

\bibitem{drs}
A. Dey, P. Roy, T. Sarkar, ArXiv : 1303.6824 [gr-qc].

\bibitem{matos}
T. Matos, F. S. Guzman, Phys. Rev. {\bf D62}, 061301 (R) (2000).

\bibitem{harako}
M.~K.~Mak, T.~Harko,  Phys.\ Rev.\ D {\bf 70}, 024010 (2004).

\bibitem{perlickreview}
V. Perlick, Living Rev. Relativity {\bf 7} (2004), 9

\bibitem{perlickconical}
V. Perlick, Phys. Rev. {\bf D69} 064017 (2004).

\bibitem{poisson}
E. Poisson, ``A Relativist's Toolkit, The Mathematics of Black-Hole Mechanics,'' Cambridge University Press (2004).

\bibitem{rising1}
W. J. G. de Blok, S. S. McGaugh,  MNRAS {\bf 290} 533 (1997).

\bibitem{rising2}
W. J. G. de Blok, S. S. McGaugh, V. C. Rubin, The Astronomical Journal {\bf 122} 2396 (2001). 

\bibitem{abell}
J. Richard, J. P. Kneib, M. Limousin, A. Edge, E. Jullo, MNRAS Letters {\bf 402}, 1, L44 (2010).

\bibitem{kochanek}
C. S. Kochanek, Astrophys. J {\bf 445} 559 (1995).

\bibitem{hewitt}
J. N. Hewitt, E. L. Turner, D. P. Schneider, B. F. Burke, G. I. Langston, C. R. Lawrence, 
Nature {\bf 333} 537 (1988).


\end{thebibliography}
\end{document}